\documentclass[epj]{svjour}
 \usepackage[dvips]{graphicx}
\begin{document}
\title{Beam-helicity asymmetry in photon and pion electroproduction in the $ \mathbf  \Delta$(1232) resonance region at $ \mathbf Q^2=$ 0.35 (GeV/c)$^2$}
\titlerunning{Beam-helicity asymmetry in photon and pion electroproduction ...}
%
%
%
%
%\author{First author\inst{1} \and Second author\inst{8}% etc
\author{
%------ first authors
I.K. Bensafa              \inst{1} \and 
%---------- then in alphabetic order
P.~Achenbach              \inst{2} \and
M.~Ases Antelo            \inst{2} \and
C.~Ayerbe                 \inst{2} \and
D.~Baumann                \inst{2} \and
R.~B\"ohm                 \inst{2} \and
D.~Bosnar                 \inst{5} \and
E.~Burtin                 \inst{3} \and 
X.~Defa\"y                \inst{1} \and 
N.~D'Hose                 \inst{3} \and 
M.~Ding                   \inst{2} \and
M.O.~Distler              \inst{2} \and
L.~Doria                  \inst{2} \and
H. Fonvieille             \inst{1}\fnmsep\thanks{{\it corresponding author,} e-mail {\tt helene@clermont.in2p3.fr}} \and 
J.M.~Friedrich            \inst{8} \and 
J.~Friedrich              \inst{2} \and 
J.~Garc{\'\i}a~Llongo     \inst{2} \and
P.~Janssens               \inst{4} \and 
G.~Jover~Ma{\~n}as        \inst{2} \and
M.~Kohl                   \inst{2} \and
G.~Laveissi\`ere          \inst{1} \and 
M.~Lloyd                  \inst{2} \and
M.~Makek                  \inst{5} \and
J.~Marroncle              \inst{3} \and 
H.~Merkel                 \inst{2} \and 
P.~Merle                  \inst{2} \and 
U.~M\"uller               \inst{2} \and
L.~Nungesser              \inst{2} \and
B.~Pasquini               \inst{7} \and 
R.~P\'erez~Benito         \inst{2} \and
J.~Pochodzalla            \inst{2} \and
M.~Potokar                \inst{6} \and
G.~Rosner                 \inst{9} \and
S.~S\'anchez Majos        \inst{2} \and 
M.~Seimetz                \inst{2,} \inst{3} \and
S.~\v{S}irca              \inst{6} \and 
T.~Spitzenberg            \inst{2} \and
G.~Tamas                  \inst{2} \and 
R.~Van~de~Vyver           \inst{4} \and 
L.~Van~Hoorebeke          \inst{4} \and 
Th.~Walcher                \inst{2} \and  
M.~Weis                   \inst{2} 
}                     % Do not remove
\institute{
%------ inst1
Laboratoire de Physique Corpusculaire IN2P3-CNRS, Universit\'e Blaise Pascal, F-63170 Aubi\`ere Cedex, France. \and
%------ inst2
Institut f\"ur Kernphysik, Johannes Gutenberg-Universit\"at, D-55099 Mainz, Germany. \and
%------ inst3
CEA Dapnia-SPhN, C.E. Saclay, F-91191 Gif-sur-Yvette Cedex, France. \and
%------ inst4
Department of Subatomic and Radiation Physics, University of Gent, B-9000 Gent, Belgium. \and
%------ inst5
Department of Physics, University of  Zagreb, HR-10002 Zagreb, Croatia. \and
%------ inst6
Institut Jo\v{z}ef Stefan, University of Ljubljana, SI-1001 Ljubljana, Slovenia. \and
%------ inst7
Dipartimento di Fisica Nucleare e Teorica, Universit\`a degli Studi di Pavia, and INFN, Sezione di Pavia, Pavia, Italy. \and
%------ inst8
Physik Department, Technische Universit\"at M\"unchen, D-85748 Garching, Germany.  \and
%------ inst9
Department of Physics and Astronomy, University of Glasgow, Glasgow G12 8QQ, UK.
} %--------- end of institutes
\date{Received: date / Revised version: date}

\abstract{
%=============================================
The beam-helicity asymmetry has been measured simultaneously for the reactions  $ \stackrel{\to}{e}\!p \to e \, p \, \gamma \, $  and  $ \stackrel{\to}{e}\!p \to e \, p \, \pi^0 \, $  in the $\Delta (1232)$ resonance region at $Q^2=$ 0.35 (GeV/c)$^2$. The experiment was performed at MAMI with a longitudinally polarized beam and an out-of-plane detection of the proton. The results are compared with calculations based on Dispersion Relations for virtual Compton scattering and with the MAID model for pion electroproduction.
There is an overall good agreement between experiment and theoretical calculations. The remaining discrepancies may be ascribed to an imperfect parametrization of some $\gamma^{(*)} N \to \pi N$ multipoles, mainly contributing to the non-resonant background. The beam-helicity asymmetry in both channels ($\gamma$ and $\pi^0$) shows a good sensitivity to these multipoles and should allow future improvement in their parametrization.
%=============================================
\PACS{
      {13.40.-f}{Electromagnetic processes and properties}   \and
      {13.60.Fz}{Elastic and Compton scattering} \and
      {13.60.Le}{Meson production} \and
      {14.20.Gk}{Baryon resonances with S=0}
     } % end of PACS codes
} %end of abstract
\maketitle

\section{Introduction}
\label{intro}

%===================================
Polarization observables are powerful tools to study hadron structure. They have seen intensive developments in the recent years in semi-inclusive and exclusive reactions, at high and low energies. At high energies they are detailed probes of mechanisms at the parton level, e.g. in the study of the transverse spin structure of the nucleon in semi-inclusive deep inelastic scattering~\cite{Ji:2006br}, or of the generalized parton distributions in exclusive deep virtual Compton scattering (DVCS)~\cite{Diehl:2003ny}.
At lower energies, models of nucleon structure have to address the non-perturbative re\-gime of QCD without the help of hard sub-mechanisms. The relevant degrees of freedom can be quarks, like in constituent quark models, or bare hadrons surrounded by a pion cloud, like in chiral perturbation theories.
Polarization observables provide specific insights, since they are more sensitive to different amplitude combinations, or interferences, than unpolarized cross sections. For example an important effort is being pursued to investigate the multipoles contributing to the formation of the $P_{33}$ resonance $\Delta(1232)$ by performing low-energy pion electroproduction experiments~\cite{Bartsch:2001ea,Elsner:2005cz}.

%===================================

Photon electroproduction on the proton, $e p \to ep \gamma$, is another interesting channel for that purpose. It gives access to virtual Compton scattering (VCS) $\gamma^* p \to \gamma p$. This process has a sensitivity to the electromagnetic structure of the nucleon that is complementary to that of, e.g., elastic scattering or pion production. 
At low energy $W$ in the $(\gamma p)$ center-of-mass system, the VCS process allows to access the generalized polarizabilities (GPs) of the nucleon, related to the VCS amplitude at vanishing energy of the outgoing photon~\cite{Guichon:1995pu}. These observables have been measured recently at different values of the four-momentum transfer squared $Q^2$ at MAMI~\cite{Roche:2000ng}, JLab~\cite{Laveissiere:2004nf} and MIT-Bates~\cite{Bourgeois:2006js}. 

\begin{figure}
\includegraphics[width=8.5cm,height=2cm]{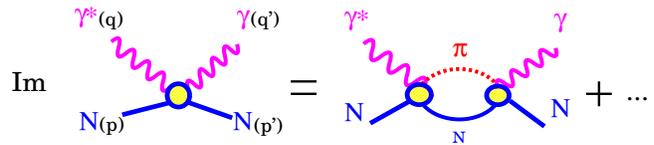}  
\caption{The imaginary part of the VCS amplitude decomposed using unitarity.}
\label{fig-imvcs}
\end{figure}

When going above the pion threshold, the VCS amplitude  $T^{VCS}$ acquires an imaginary part due to the coupling to the $\pi N$ channel. Therefore single polarization observables, which are proportional to $\mathcal{I}m (T^{VCS})$, become non-zero above pion threshold. A particularly relevant observable is the single-spin asymmetry (SSA) or beam-helicity asymmetry:
%----------- 
$SSA =(\sigma ^{\uparrow}  -  \sigma ^{\downarrow}) / 
 (\sigma ^{\uparrow}  +  \sigma ^{\downarrow}) $ where $\sigma ^{\uparrow}$ and $\sigma ^{\downarrow}$ designate the photon electroproduction cross section with beam helicity state + and $-$, respectively.
%-----------
  As it was first pointed out in ref.~\cite{Kroll:1995pv}, the SSA yields direct information on the absorptive part of the VCS amplitude, and on the relative phase between the VCS amplitude and the Bethe-Heitler (BH) contribution. The BH process refers to the photon emission by the incoming or outgoing electron and it adds coherently to the VCS amplitude. Moreover, the VCS amplitude can be split into a Born part, given in terms of nucleon ground state properties (the electromagnetic form factors), and a non-Born part which contains all nucleon excitations and meson-loop contributions. Since the BH and Born-VCS contributions are purely real, the SSA is proportional to the imaginary part of the non-Born VCS amplitude. In particular, the numerator of the SSA can be written as $\mathcal{I}m (T^{VCS})\cdot \mathcal{R}e (T^{VCS}+T^{BH})$. After development, one obtains the sum of a pure VCS contribution and a VCS-BH interference term which has the effect to enhance the asymmetry. The absorptive part of the VCS amplitude can be obtained, through unitarity relation, from the photo- and electro-production amplitudes on the nucleon. In the region of $W\sim 1.2$ GeV, the most important contribution is from $\pi N$ intermediate states, as schematically depicted in fig.~\ref{fig-imvcs}, while mechanisms involving more pions or heavier mesons in the intermediate states are suppressed. Regarding the $Q^2$-dependence, the pion photoproduction description is on solid experimental grounds, while electroproduction data are scarce. Therefore a measurement of the beam SSA in the $\Delta$-resonance region gives a direct test of how well the description of the VCS amplitude holds, in terms of the available phenomenological information on pion photo- and electro-production amplitudes. This is the main purpose of the experiment described in the present paper.

In the case of  DVCS one has a known $\phi$-dependence of the numerator and denominator of the SSA. In our kinematic regime this dependence is not known analytically. However, the Dispersion Relations (DR) model calculation discussed in section~\ref{sec-discuss} gives a shape of the asymmetry close to a $\sin \phi$, despite the distortion of the numerator and denominator due to the BH process.
%--------

Since in the experiment the reaction $e p \to ep \pi^0$ was detected too, the beam SSA was also measured in this channel, complementing previous measurements of this type at different kinematics~\cite{Bartsch:2001ea,Joo:2003uc}. In pion electroproduction, the beam SSA, also called $\rho_{LT}'$, is proportional to the fifth structure function $R_{LT}'$~\cite{Drechsel:1992pn} and is mainly sensitive to the multipole ratios $S_{1+}/M_{1+}$ and $S_{0+}/M_{1+}$ in the region of the $\Delta$ resonance.

%--------------------------------------------------------
%------------------------------------------------------------

%-----------------------
%-----------------------
%-----------------------

\section{The experiment}
\label{sec:1}

 The experiment was performed at the Mainz Microtron MAMI using the 100 \% duty cycle electron beam at an energy of 883.2 MeV, allowing for a longitudinal beam polarization $P_e$ of 75-85 \%. Helicity reversal was performed every second, and beam current values were typically 13-25 $\mu A$. The experiment used the standard equipment of the A1 collaboration~\cite{Blomqvist:1998xn}: the M\o{}ller polarimeter to measure $P_e$ once a day, the 5 cm long liquid Hydrogen target, spectrometer A to detect  the scattered electron at a fixed setting, and spectrometer B to detect the final proton at three different out-of-plane settings. Table 1 summarizes the kinematical settings, corresponding to an average $Q^2$ of 0.35 (GeV/c)$^2$ and a virtual photon momentum in the center-of-mass $q_{cm}=600$ MeV/c, similar to a previous VCS experiment at MAMI~\cite{Roche:2000ng}. However, here $W$ is above pion threshold ($W \sim$ 1.2 GeV) and the virtual photon polarization $\epsilon=0.48$ is the highest achievable at this value of $W$.

%-------------------------------------------------
\begin{table}
\caption{Experimental settings for spectrometers A (electron arm) and B (proton arm): values of the horizontal angles $\theta_A$ and $\theta_B$  w.r.t. the beam direction, and the central momenta $p_A$ and $p_B$. $\alpha_{oop}$ is the out-of-plane angle of spectrometer B.
}
\label{tabhf1} 
\begin{tabular}{cccccc}
\noalign{\smallskip}\hline\noalign{\smallskip}
 $\theta_A$  & $p_A$  &    \ \ $\theta_B$   & $p_B$  &  $\alpha_{oop}$    & \ \ setting \\
    &  (MeV/c)  & \ &  (MeV/c) & \  \\
\noalign{\smallskip}\hline\noalign{\smallskip}
 \ & \ &    \ \ 25.2$^{\circ}$  &    345  &  2$^{\circ}$   &  \ \ OOP-1 \\
  59.9$^{\circ}$ & 401.2  & \ \ 20.3$^{\circ}$  & 358  &  7$^{\circ}$   & \ \ OOP-2 \\
 \ &  \ & \ \ 15.0$^{\circ}$  &  398  &  10$^{\circ}$   & \ \ OOP-3 \\
\noalign{\smallskip}\hline
\end{tabular}
\end{table}
%-------------------------------------------------

The detector package in each spectrometer includes a set of two double-planes of vertical drift chambers (VDC) for particle tracking and two segmented planes of scintillators for particle identification and timing measurements. The experiment also uses the threshold gas \v{C}erenkov coun\-ter in spectrometer A for electron identification. The beam was off-centered horizontally in order to minimize the pathlength in Hydrogen for the low-energy emitted protons.
%---------------------
Analysis cuts include particle identification for the final electron and proton, good quality tracks in the VDCs, and rejection of backscattered protons at the entrance window of spectrometer B. A specific cut is applied to eliminate protons emitted at the most upward angles, which hit a piece of the target holding system.
%-------------------
Empty target cell studies showed that the rate of background events from $ e A \to e  p  X $ reactions in the target walls was negligible compared to the rate of $ep \to ep \pi^0$ events, but not negligible compared to the rate of $ep \to ep \gamma$ events. Therefore a cut on the target length is applied in the analysis of the $\gamma$ channel, but not in the $\pi^0$ channel.

%--------------------------------------------------------
\begin{figure}
\includegraphics[width=8.0cm,height=6.5cm]{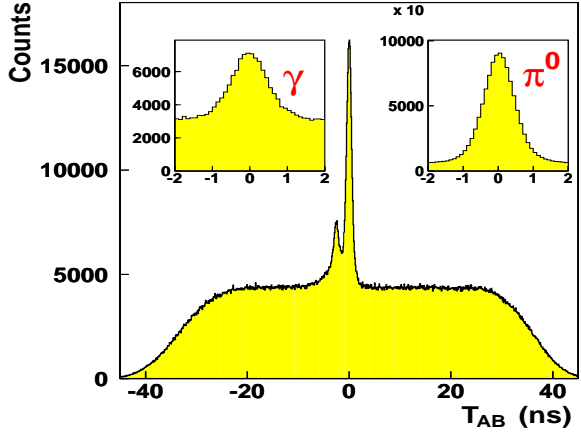}
\caption{The coincidence time spectrum for a sample of the data (no cuts). Inserts: zoom on the central peak after analysis cuts, for the full statistics in both channels: $\gamma$ (left) and $\pi^0$ (right) electroproduction.}
\label{fig-tab}
\end{figure}
%------------------------------------------------------------

Fig.~\ref{fig-tab} shows the spectrum of coincidence time ($T_{AB}$) for triggers in coincidence between the two spectrometers. The central peak represents the true $(e,p)$ coincidences and has a FWHM of 1 ns. Without any cuts, one notices a secondary peak near $T_{AB}=-2$ ns due to true $(\pi^-,p)$ coincidences, which is eliminated by a cut on the \v{C}erenkov counter signal in spectrometer A. After all cuts, the level of random coincidences is still high for the $\gamma$ channel, as illustrated by the insert on fig.~\ref{fig-tab}. Events are selected  within $\pm$ 1.4 ns in the central peak, and random coincidences are subtracted using side-band events.

\begin{figure}
\includegraphics[width=8.5cm,height=8.0cm]
{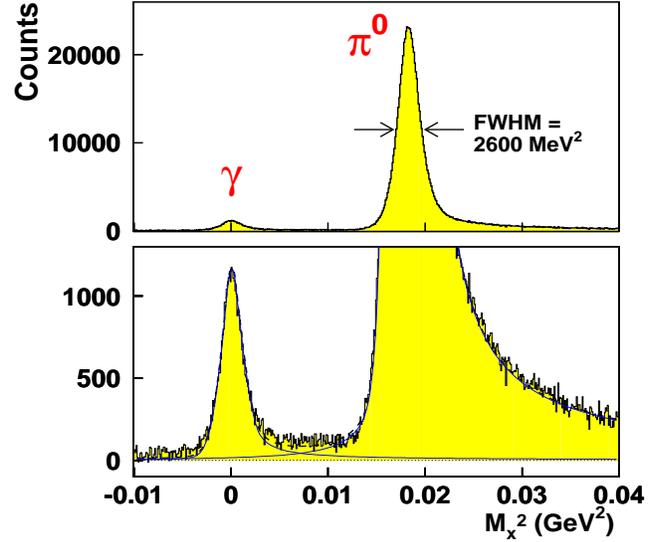}
\caption{Top: the missing mass squared spectrum within analysis cuts and after subtraction of random coincidences. Bottom: the same histogram truncated in ordinate to enhance the photon peak, plus a fit function drawn for each peak (solid lines) and the sum (dashed line).}
\label{fig-mx2}
\end{figure}

The four-momentum of the third, undetected particle is reconstructed by the missing energy and missing momentum associated to the detected $(ep)$ system. Fig.~\ref{fig-mx2} shows the final spectrum of the missing mass squared ($M_x^2$). The most prominent peak is due to $\pi^0$ electroproduction events. The second peak, centered on $M_x^2=0$, is due to photon electroproduction events, and its smallness reflects the smallness of the corresponding cross section.  This spectrum is obtained after a careful calibration of experimental parameters, such as the beam position along the horizontal transverse axis, a casual layer of frost on the target walls, and the central momentum of spectrometer B for each run period, which was changed by less than a few 10$^{-3}$ w.r.t. its nominal value. The separation between the two peaks in $M_x^2$  is well-achieved. Both peaks are similar in shape: they have the same central width due to experimental resolution, and a radiative tail extending to larger values. An empirical fit of this shape (see curves on fig.~\ref{fig-mx2}) allows to quantitatively estimate the residual $\pi^0$ events under the photon peak, yielding a contamination $C=4$~\%. 
%----------
All other background processes, e.g. coming from the target walls,  are reduced to a negligible level within the analysis cuts.
%----------
Finally, the two physics channels are selected by the following cuts in $M_x^2$: [$-$0.005, 0.005] GeV$^2$ for the $ep \to ep \gamma$  process and [0.013, 0.029] GeV$^2$ for the $ep \to ep \pi^0$ process. The achieved statistics are then 38k true $\gamma$ events and 1M true $\pi^0$ events. More details on the analysis  can be found in refs.~\cite{Bensafa:2006,vcs-ssa-1}.

%--------------------------------------------------------
%------------------------------------------------------------

%=============================
%=============================
%=============================
\section{Analysis method and results}
\label{sec:2}

\begin{figure}
\centering\includegraphics[width=7.0cm,height=6.5cm]{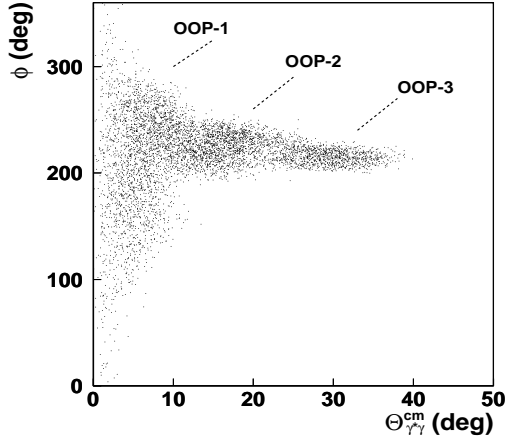}
\caption{The accepted phase space in terms of the polar and azimuthal angles of the Compton scattering process.}
\label{fig-thetaphi}
\end{figure}

\begin{figure}
\centering\includegraphics[width=7.5cm,height=2.3cm]{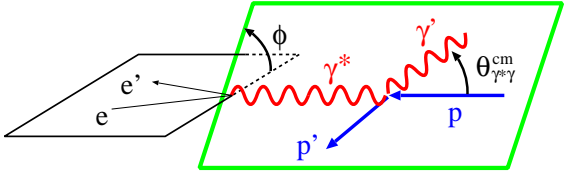}
\caption{Kinematics of the $ep \to ep \gamma$ reaction in the $\gamma p$ center-of-mass.}
\label{fig-kin}
\end{figure}

Fig.~\ref{fig-thetaphi} shows the angular phase space coverage for photon electroproduction events: $\theta_{\gamma^* \gamma}^{cm}$ is the polar angle between the initial and final photons of the Compton scattering process and $\phi$ is the azimuthal angle between the leptonic and hadronic planes as illustrated on figure~\ref{fig-kin}. The value of $\phi=+90^{\circ}$ corresponds to the missing particle emitted along the direction of $\vec k \times \vec k'$, where $\vec k$ and  $\vec k'$ are the momenta of the incoming and scattered electrons. The three settings cover altogether the region of forward polar angles up to 40$^{\circ}$, and a domain in $\phi$ narrowing around 220$^{\circ}$ as $\theta_{\gamma^* \gamma}^{cm}$ increases. According to model calculations, the asymmetry is maximal at forward  polar angles for VCS, whereas it is maximal at backward angles for $\pi^0$ electroproduction. The region of forward angles was chosen, since the experiment was optimized for VCS.

%--------------------------------------------------------
%------------------------------------------------------------

%--------------------------------------------------------
%------------------------------------------------------------

Since the SSA is a relative quantity, the data do not need to be corrected for detector inefficiencies, deadtimes, etc. 
%-------------
Only false asymmetries were checked, and found to be very small: the beam charge asymmetry (for which the data are corrected) is of the order of $10^{-3}$.
%-------------
The phase space is binned in $\theta_{\gamma^* \gamma}^{cm}$ (4$^{\circ}$ wide, large compared to the experimental resolution) and in each $\theta$-bin the SSA is determined as one value, corresponding to one central kinematical point. To this end the asymmetry is fitted to the assumption $ SSA= K \cdot \sin \phi$ (cf. section~\ref{intro}) and the  factor $K$ is determined by a maximum likelihood method. The probability associated to event $i$ is proportional to $(1 +  h_i P_e K \sin \phi_i)$ where $h_i$ is the beam helicity state (+1 or $-$1) and $\phi_i$ the azimuthal angle of the event. The likelihood method includes the treatment of random coincidences and also yields the statistical error on $K$. As a cross-check, a more classical method is used, based on the count rates $N^+$ and $N^-$: the asymmetry $A=(N^+ - N^-)/(N^+ + N^-) /P_e$ is calculated in finite bins in $\phi$ and then fitted to the same shape $A= K \cdot \sin \phi$.

The final asymmetry is computed as $K \cdot \sin 220^{\circ}$, i.e. it is projected to $\phi=220^{\circ}$, a value close to the average over the acceptance. Since the central kinematics vary slightly from bin to bin, one makes a projection to a fixed, nominal kinematics in $(Q^2, W, \epsilon )$ :
%*********************************
\begin{eqnarray*}
SSA^{exp}_{nom.kin.} = SSA^{exp}_{exp.kin.} \ \times \ {\displaystyle  SSA^{model}_{nom.kin.} \over \displaystyle SSA^{model}_{exp.kin.} }
\label{eq-proj} 
\end{eqnarray*}
%*********************************
by use of a model: Dispersion Relations~\cite{Pasquini:2001yy} for $ep \to ep \gamma$ and the unitary isobar model  MAID~\cite{Drechsel:1998hk} for $ep \to ep \pi^0$.
%-----------------
The nominal point corresponds to our average $ep \to ep \gamma$  kinematics: $Q^2=0.35$ (GeV/c)$^2$, $W=1.19$ GeV and $\epsilon=0.48$. For the $\pi^0$ channel, the accepted phase space is very similar. The same analysis method is used, and the asymmetry is projected from the experimental average ($Q^2=0.34$ (GeV/c)$^2$, $W=1.21$ GeV, $\epsilon=0.47$) to the same nominal kinematics as above. This choice results in all projection factors as close as possible to 1, which is desirable in order to minimize projection errors.

Systematic errors are estimated from the sources of uncertainty listed in table~\ref{tabhf2}. These errors are propagated to the asymmetry, either analytically (items 1 to 4) or by running the analysis in  different conditions (items 5, 6) and taking the spread of the resulting SSA. Concerning item 2, the SSA in photon electroproduction is corrected for $\pi^0$ contamination, using $SSA = SSA_{raw} \cdot (1+C)$ to first order, and a $\pm$ 50\% uncertainty is applied on  $C$. Concerning item 3, radiative corrections to the asymmetry are generally small, as shown in~\cite{Vanderhaeghen:2000ws,Afanasev:2005pb}. They have been computed for the present kinematics~\cite{vcs-ssa-2,Smirnov:2005vh} and found negligible, therefore they are ignored in the analysis. However, an error is assigned to this procedure; the error considered in table 2, item 3,  is a way to account for model uncertainties that are present in the calculation of radiative corrections. Items 1 to 4 of this table yield very small errors; the dominant systematic errors come from items 5 and 6. The main errors of the calibration are in the electron scattering angle and the proton-arm central momentum; other calibration parameters (see section~\ref{sec:1}) have much less influence. The quadratic sum of partial systematic errors bin-per-bin  is presented in table~\ref{tabhf3}  together with our final SSA result and its statistical error. The latter is still by far the dominant error. In the VCS channel, the statistical error is enlarged by a factor $\sim$ 1.8 due to the high level of random coincidences.
%-------------------------------------------------
\begin{table}
\caption{Summary of systematic errors (more details can be found in~\cite{vcs-ssa-1}).}
\label{tabhf2} 
\begin{tabular}{ll}
\noalign{\smallskip}\hline\hline\noalign{\smallskip}
type of error   & amount of error  \\
\noalign{\smallskip}\hline\hline\noalign{\smallskip}
1) beam polarization     &   $\Delta P_e/P_e = \pm 0.02$ \\
\noalign{\smallskip}\hline\noalign{\smallskip}
2) $\pi^0$ contamin. ($\gamma$)  & $\Delta C= \pm$ 2 \% \ ($C=$ 4 \%) \\
\noalign{\smallskip}\hline\noalign{\smallskip}
3) radiative correction & $\pm$ 100 \% of the correction \\
 \noalign{\smallskip}\hline\noalign{\smallskip}
4) proj. to nom.kin. & proj. factor varied by $\pm$ 10 \% \\ 
\noalign{\smallskip}\hline\noalign{\smallskip}
5) calibration of  & $\Delta \theta_A = \pm 0.5$ mrad (electron) \\
\ \ \ momenta and angles  & $\Delta p_B/p_B = \pm 1 \cdot 10^{-3}$ 
(proton) \\
\noalign{\smallskip}\hline\noalign{\smallskip}
6) stability within cuts & cuts in $T_{AB}$, acceptance, $M_x^2$ etc. \\ 
\noalign{\smallskip}\hline
\end{tabular}
\end{table}
%-------------------------------------------------

%==========================================================
\begin{table}
\caption{Beam SSA result for the two reactions, at nominal kinematics: \ $Q^2$=0.35 (GeV/c)$^2$, $W$=1.19 GeV, $\epsilon$=0.48, $\phi=220^{\circ}$.}
\label{tabhf3}  
\begin{tabular}{lccc}
\hline\hline\noalign{\smallskip}
\multicolumn{4}{c}{ $ \stackrel{\to}{e} p \to e \ p \ \gamma$ \ channel} \\
\noalign{\smallskip}\hline\noalign{\smallskip}
$\theta_{\gamma^* \gamma}^{cm}  (^{\circ})$ & \ \ \ \ \ \ \ \ $SSA$ \ \ \ \ \ \ \ & \ $\Delta SSA_{stat}$ & \ $\Delta SSA_{syst}$ \\
\noalign{\smallskip}\hline\noalign{\smallskip}
    2.6  &    -0.021   &    0.026   &    0.015    \\   
    6.0  &    -0.035   &    0.017   &    0.006    \\
    9.8  &    -0.032   &    0.020   &    0.011    \\
   14.0  &    -0.026   &    0.030   &    0.016    \\
   17.9  &    -0.014   &    0.034   &    0.020    \\
   21.9  &    -0.121   &    0.044   &    0.021    \\
   26.1  &    -0.050   &    0.046   &    0.023    \\
   29.9  &    -0.053   &    0.050   &    0.034    \\
   33.7  &    -0.004   &    0.070   &    0.047    \\
\noalign{\smallskip}\hline\hline\noalign{\smallskip}
\multicolumn{4}{c}{ $ \stackrel{\to}{e} p \to e \ p \ \pi^0$ \ channel} \\
\noalign{\smallskip}\hline\noalign{\smallskip}
$\theta_{\gamma^* \pi}^{cm}  (^{\circ})$ & $SSA$  &  $\Delta SSA_{stat}$ &  $\Delta SSA_{syst}$ \\
\noalign{\smallskip}\hline\noalign{\smallskip}
    2.5  &     0.0016   &    0.0035   &    0.0017    \\
    5.9  &    -0.0007   &    0.0028   &    0.0011    \\
    9.9  &    -0.0059   &    0.0033   &    0.0024    \\
   14.0  &    -0.0064   &    0.0040   &    0.0018    \\
   17.9  &    -0.0077   &    0.0043   &    0.0021    \\
   22.1  &    -0.0055   &    0.0053   &    0.0031    \\
   26.2  &    -0.0033   &    0.0043   &    0.0022    \\
   30.0  &    -0.0032   &    0.0039   &    0.0016    \\
   33.7  &    -0.0078   &    0.0049   &    0.0014    \\
   37.4  &    -0.0205   &    0.0090   &    0.0039    \\
\noalign{\smallskip}\hline
\end{tabular}
\end{table}
%================================================

%=======================
%=======================
%=======================
\section{Discussion} \label{sec-discuss}

%--------------------------------------------------------
\begin{figure}
\includegraphics[width=8.0cm,height=7.5cm]{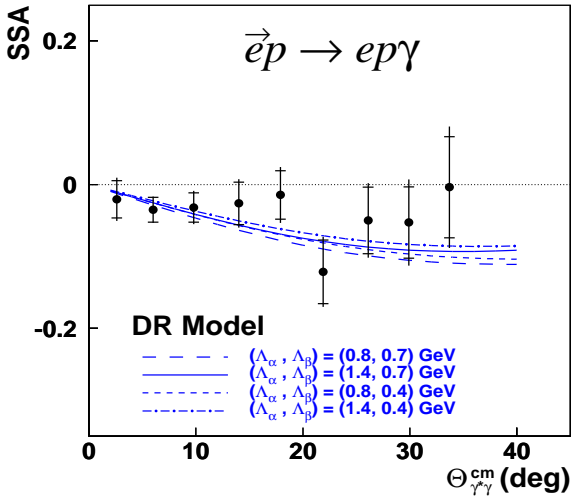}
\caption{The beam SSA in photon electroproduction  at nominal kinematics compared to the calculation of the DR model (using MAID2003 multipoles) for several values of the free parameters $\Lambda_{\alpha}, \Lambda_{\beta}$. The inner error bar is statistical, the outer one is the quadratic sum of statistical and systematic errors. }
\label{fig-res-vcs}
\end{figure}
%------------------------------------------------------------

\subsection{Photon electroproduction channel} \label{sec-discuss-gam}

The result of the measurement of $\stackrel{\to}{e}p \to ep \gamma$ is displayed on fig.~\ref{fig-res-vcs} together with the calculation of the DR formalism of refs.~\cite{Pasquini:2001yy,Drechsel:2002ar}. This model has been developed for real and virtual Compton scattering at moderate energies up to the $\pi \pi N$ threshold. It is the only model available in the literature that calculates specifically $\mathcal{I}m (T^{VCS})$, and therefore is able to predict the beam SSA. The VCS invariant amplitudes are given by $s$-channel dispersion integrals, in which the imaginary part is calculated through unitarity, taking into account the contribution from $\pi N$ intermediate states. Multipoles for $\gamma^{(*)} N \to \pi N$ are  taken from the MAID 2003 analysis. 

The DR model has two free parameters, $\Lambda_{\alpha}$ and $\Lambda_{\beta}$, related to the electric and magnetic GPs and parametrizing their unconstrained part, i.e. asymptotic behaviour and dispersive contribution beyond $\pi N$. It is clear from fig.~\ref{fig-res-vcs} that the SSA exhibits little sensitivity to the GPs. By comparison to the theoretical curves displayed on the figure, one can state that for this given set of $\pi N$ multipoles the trend of the data is to favor high values of $\Lambda_{\alpha}$ and low values of  $\Lambda_{\beta}$, i.e. high values of the electric GP and high values of the magnetic GP. As a reminder, in the DR model the electric GP increases with  $\Lambda_{\alpha}$, while the magnetic GP decreases with  $\Lambda_{\beta}$.

%--------------------------------------------------------
\begin{figure}
\includegraphics[width=8.0cm,height=7.5cm]{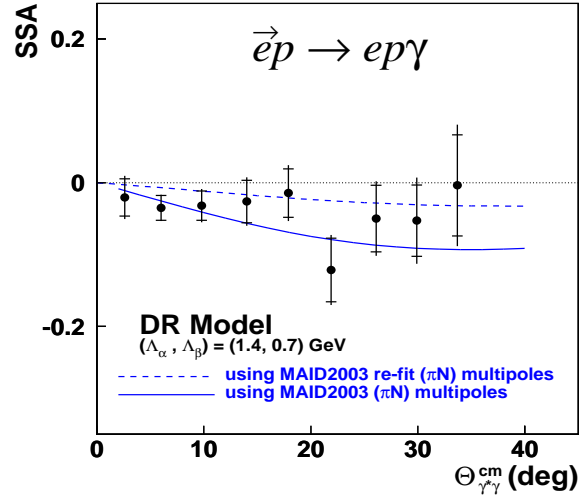}
\caption{The beam SSA in photon electroproduction: sensitivity to the MAID $(\pi N)$ multipoles. }
\label{fig-res-vcs2}
\end{figure}
%------------------------------------------------------------

As explained above, the main feature of the beam SSA is that it is fully sensitive to $\mathcal{I}m (T^{VCS})$. Therefore our measurement is a direct test of the calculation of this quantity. The experimental shape of the SSA in fig.~\ref{fig-res-vcs} is rather smooth, similarly to the DR  prediction. The measured asymmetry is in overall good agreement with the theoretical calculation, although the latter tends to overestimate the magnitude of the asymmetry gradually as $\theta_{\gamma^* \gamma}^{cm}$ increases. This suggests that the calculation of $\mathcal{I}m (T^{VCS})$ can be improved. In that respect, the  $\pi N$ multipoles, which are an essential input to the DR calculation of VCS, may need a better parametrization. It is worth noting that on the one hand the beam SSA in VCS as given by the DR model is insensitive to the change of parametrization from the ``2000'' to  ``2003'' version of MAID (at least in our kinematics). On the other hand it is sensitive to further ``re-fits'' such as the one cited in ref.~\cite{Elsner:2005cz}, which involves an adjustment of the longitudinal multipoles $S_{1+}$ and $S_{0+}$ in the $\pi^0 p$ channel. In particular, figure~\ref{fig-res-vcs2} shows that the calculation with this re-fit gives a lower asymmetry, in better agreement with the trend of the data at increasing angles. Nevertheless, since this MAID re-fit~\cite{Elsner:2005cz} was performed at kinematics different from the present experiment, there is room left for further $\pi N$ multipole adjustment. Other possibilities like contributions beyond $\pi N$ are very unlikely, since our measured range in $W$ is mostly below the two-pion threshold.

%--------------------------------------------------------
\begin{figure}
\includegraphics[width=8.0cm,height=7.5cm]{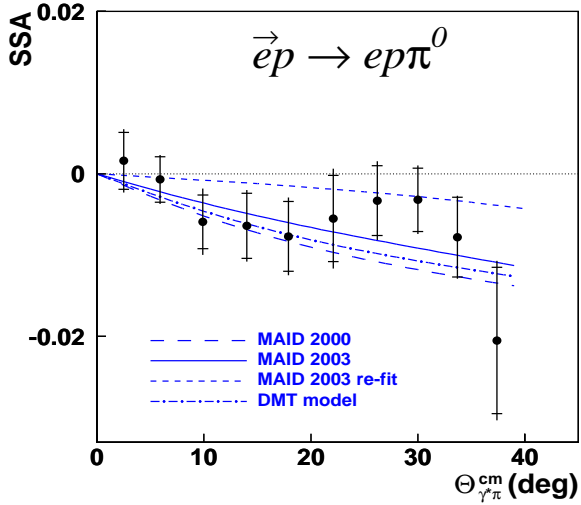}
\caption{The beam SSA in $\pi^0$ electroproduction  at nominal kinematics, compared to theoretical calculations. Same convention as figure~\ref{fig-res-vcs} for the error bars.}
\label{fig-res-piz}
\end{figure}
%------------------------------------------------------------

\subsection{Pion electroproduction channel} \label{sec-discuss-piz}

The result of the measurement of $\stackrel{\to}{e}p \to ep \pi^0$ is displayed on fig.~\ref{fig-res-piz}. At small angles up to $\theta_{\gamma^* \pi}^{cm}=20^{\circ}$, the data are in very good agreement with the MAID model~\cite{Drechsel:1998hk} (versions ``2000'' and ``2003'') and the DMT model~\cite{Kamalov:1999hs}.
Also shown is the MAID re-fit~\cite{Elsner:2005cz} mentioned above, in which the longitudinal multipoles $S_{1+}$ and $S_{0+}$ have been adjusted essentially at $Q^2$=0.2 (GeV/c)$^2$. Applied to our $Q^2$ of 0.35 (GeV/c)$^2$~\cite{lothar2006}, this version yields a much lower asymmetry than the other predictions, and not in better agreement with our measurement. It was already suggested in \cite{Joo:2003uc} that a simple rescaling of the $S_{1+}$ and $S_{0+}$ multipoles cannot account for the $Q^2$-dependence of the data.

%-----------
Beyond 20$^{\circ}$ the data of fig.~\ref{fig-res-piz} tend to suggest some structure which no model can reproduce. In the $s$ and $p$ wave $M_{1+}$-dominance approximation, the fifth structure function $R_{LT}'$ is proportional to \ 
$\sin \theta_{\gamma^* \pi}^{cm} \cdot \mathcal{I}m \{ (6 \cos \theta_{\gamma^* \pi}^{cm} S_{1+}$ + $ S_{0+} )^* M_{1+} \} $ \cite{Drechsel:1992pn}.
To describe the rapid $\theta$-variations that we observe for the SSA is quite challenging; it requires the contribution from higher-order multipoles which produce a non-resonant background modulating the resonant $M_{1+}$  contribution, as already noticed in ref.~\cite{Joo:2003uc}.

%====================
%====================
%====================
%==================== conclusion
\section{Conclusion}

The beam-helicity asymmetry has been measured simultaneously for photon and $\pi^0$ electroproduction in the $\Delta$(1232) resonance region. The measured asymmetries are of the order of a few percent for $\gamma$ and smaller than 1\% for $\pi^0$. There is an overall good agreement between our measurement and the theoretical calculations, based on the DR and MAID models. The remaining discrepancies in shape or magnitude of the asymmetry might be ascribed to an imperfect parametrization of some $\pi N$ multipoles, mainly contributing to the non-resonant background. 

Concerning virtual Compton scattering, which was the main goal of the experiment, we have performed the first measurement of the beam SSA in photon electroproduction at low energies. The latter provides an important cross-check for the input to the DR formalism for VCS since the imaginary part  of the VCS amplitude is connected through unitarity to the $\gamma^{(*)} N \to \pi N$ multipoles. In order to improve the agreement between experiment and theory for the beam SSA in the photon and electroproduction channels, one could attempt a simultaneous fit of the two observables, by changing the parametrization of some $\pi^0$ electroproduction multipoles. In that view, the observables in the two physics channels become coupled. Therefore the data presented in this paper address in a new way important questions by showing how the simultaneous measurements in several de-excitation channels ($\gamma N$ and $\pi N$) can help to gain new insights for our understanding of the nucleon and resonance phenomena at low energy.

%----- thanks
\vskip 3 mm
We thank the MAMI accelerator staff for providing the excellent polarized beam. We thank L.Tiator for valuable discussions and for providing his MAID re-fit result, and M.Vanderhaeghen and G.Smirnov for their contribution to the calculation of radiative corrections. This work was supported by the Deutsche Forschungsgemeinschaft (SFB 443), the Federal State of Rhineland-Palatinate, the French CEA and CNRS/IN2P3, the BOF-Gent University and FWO-Flanders (Belgium) and the EPSRC, UK.

% =====================================

\bibliographystyle{h-elsevier2.bst}

\bibliography{common}

\begin{thebibliography}{10}

\bibitem{Ji:2006br}
X. Ji et~al.,
\newblock Phys. Lett. B638 (2006) 178, hep-ph/0604128.

\bibitem{Diehl:2003ny}
M. Diehl,
\newblock Phys. Rept. 388 (2003) 41, hep-ph/0307382.

\bibitem{Bartsch:2001ea}
P. Bartsch et~al.,
\newblock Phys. Rev. Lett. 88 (2002) 142001, nucl-ex/0112009.

\bibitem{Elsner:2005cz}
D. Elsner et~al.,
\newblock Eur. Phys. J. A27 (2006) 91, nucl-ex/0507014.

\bibitem{Guichon:1995pu}
P.A.M. Guichon, G.Q. Liu and A.W. Thomas,
\newblock Nucl. Phys. A591 (1995) 606, nucl-th/9605031.

\bibitem{Roche:2000ng}
J. Roche et~al.,
\newblock Phys. Rev. Lett. 85 (2000) 708.

\bibitem{Laveissiere:2004nf}
Jefferson Lab Hall A, G. Laveissiere et~al.,
\newblock Phys. Rev. Lett. 93 (2004) 122001, hep-ph/0404243.

\bibitem{Bourgeois:2006js}
P. Bourgeois et~al.,
\newblock Phys. Rev. Lett. 97 (2006) 212001, nucl-ex/0605009.

\bibitem{Kroll:1995pv}
P. Kroll, M. Schurmann and P.A.M. Guichon,
\newblock Nucl. Phys. A598 (1996) 435, hep-ph/9507298.

\bibitem{Joo:2003uc}
CLAS, K. Joo et~al.,
\newblock Phys. Rev. C68 (2003) 032201, nucl-ex/0301012.

\bibitem{Drechsel:1992pn}
D. Drechsel and L. Tiator,
\newblock J. Phys. G18 (1992) 449.

\bibitem{Blomqvist:1998xn}
K.I. Blomqvist et~al.,
\newblock Nucl. Instrum. Meth. A403 (1998) 263.

\bibitem{Bensafa:2006}
I.K. Bensafa,
\newblock PhD thesis, UBP Clermont-Ferrand, 2006,
\newblock {D}U 1647.

\bibitem{vcs-ssa-1}
. H.Fonvieille and . I.K.Bensafa,
\newblock LPC Internal reports PCCF-RI-0604 and PCCF-RI-0605 (2006).

\bibitem{Pasquini:2001yy}
B. Pasquini et~al.,
\newblock Eur. Phys. J. A11 (2001) 185, hep-ph/0102335.

\bibitem{Drechsel:1998hk}
D. Drechsel et~al.,
\newblock Nucl. Phys. A645 (1999) 145, nucl-th/9807001,
\newblock http://www.kph.uni-mainz.de/MAID/.

\bibitem{Vanderhaeghen:2000ws}
M. Vanderhaeghen et~al.,
\newblock Phys. Rev. C62 (2000) 025501, hep-ph/0001100.

\bibitem{Afanasev:2005pb}
A.V. Afanasev, M.I. Konchatnij and N.P. Merenkov,
\newblock J. Exp. Theor. Phys. 102 (2006) 220, hep-ph/0507059.

\bibitem{vcs-ssa-2}
. H.Fonvieille,
\newblock LPC Internal report PCCF-RI-0607 (2006).

\bibitem{Smirnov:2005vh}
G.I. Smirnov,
\newblock (2005), hep-ph/0504045.

\bibitem{Drechsel:2002ar}
D. Drechsel, B. Pasquini and M. Vanderhaeghen,
\newblock Phys. Rept. 378 (2003) 99, hep-ph/0212124.

\bibitem{Kamalov:1999hs}
S.S. Kamalov and S.N. Yang,
\newblock Phys. Rev. Lett. 83 (1999) 4494, nucl-th/9904072,
\newblock http://www.kph.uni-mainz.de/MAID/dmt/.

\bibitem{lothar2006}
. L.Tiator,
\newblock private communication.

\end{thebibliography}
% =====================================

\end{document}